\documentclass[authoryear,12pt]{elsarticle}

\usepackage{url}
\usepackage{graphicx}        % standard LaTeX graphics tool
\usepackage{subfigure}
\usepackage{amsmath}
\usepackage{amssymb}
\usepackage{amsthm}
\usepackage[bottom]{footmisc}% places footnotes at page bottom

\journal{...}

\begin{document}

\begin{frontmatter}

\title{Why Global Performance is a Poor Metric for Verifying Convergence of Multi-agent Learning}

\author[buid,edin]{Sherief Abdallah\corref{sa}}
\address[buid]{British University in Dubai, United Aran Emirates} 
\address[edin]{University of Edinburgh, United Kingdom}
\cortext[sa]{Email: shario@ieee.org}

\begin{abstract}
Experimental verification has been the method of choice for verifying the stability of a multi-agent reinforcement learning (MARL) algorithm as the number of agents grows and theoretical analysis becomes prohibitively complex. 
For cooperative  agents, where the ultimate goal is to optimize some global metric, the stability is usually verified by observing the evolution of the global performance metric over time. If the global metric improves and eventually stabilizes, it is considered a reasonable verification of the system's stability. 

The main contribution of this note is establishing the need for better experimental frameworks and measures to assess the stability of large-scale adaptive cooperative systems.
We show an experimental case study where the stability of the global performance metric can be rather deceiving, hiding an underlying instability in the system that later leads to a significant drop in performance. We then propose an alternative metric that relies on  agents' local policies and show, experimentally, that our proposed metric is more effective (than the traditional global performance metric) in exposing the instability of MARL algorithms. 
\end{abstract}


\end{frontmatter}

%\linenumbers

%------------------------------------------------------------------------- 
\section{Introduction}
\label{sec-intro}

The term \emph{convergence}, in reinforcement learning context, refers to the stability of the learning process (and the underlying model) over time. 
Similar to single agent reinforcement learning algorithms (such as Q-learning \cite{sutton:rl}), the convergence of a multi-agent reinforcement learning (MARL) algorithm is an important property that received considerable attention \cite{bowling02aij,bowling05nips,conitzer07mlj,sherief08jair}. 
However, proving the convergence of a MARL algorithm via theoretical analysis is significantly more challenging than proving the convergence in the single agent case. The presence of other agents that are also learning deem the environment non-stationary, therefore violating a foundational assumption in single agent learning. In fact, proving the convergence of MARL algorithm even in 2-player-2-action single-stage games (arguably the simplest class of multi-agent systems domains) has been challenging \cite{bowling02aij,conitzer07mlj,sherief08jair}. 

As a consequence, experimental verification is usually the method of choice as the number of agents grows and theoretical analysis becomes prohibitively complex. 
For cooperative agents, researchers typically verified  the stability of a MARL algorithm by observing the evolution of some global performance metric overtime \cite{boyan94nips,peshkin02ijcnn,mohamad06jaamas,sherief07aamas, sherief08jair}. This is not surprising since the ultimate goal of a cooperative system is to optimize some global metric. Examples of global performance metrics include the percentage of total number of delivered packets in routing problems \cite{chang04icac}, the average turn around time of tasks in task allocation problems  \cite{sherief08jair}, or the average reward (received by agents) in general \cite{mohamad06jaamas}. 

If the global metric improves over time and eventually \emph{appears} to stabilize, it is usually considered a reasonable verification of convergence \cite{boyan94nips,peshkin02ijcnn,mohamad06jaamas,sherief07aamas, sherief08jair}. Even if the underlying agent policies are not stable, one could argue that at the end, global performance is all that matters in a cooperative system.

This paper challenges the above (widely-used) practice and establishes the need for better experimental frameworks and measures for assessing the stability of large-scale cooperative systems.
We show an experimental case study where the stability of the global performance metric can hide an underlying instability in the system. This hidden instability later leads to a significant drop in the global performance metric itself. We propose an alternative measure that relies on agents' local policies: the policy entropy. We experimentally show that the proposed metric is more effective than the traditional global performance metric in exposing the instability of MARL algorithms in large-scale multi-agent systems. 

The paper is organized as follows. 
Section \ref{sec-case} describes the case study we will be using throughout the paper.
Section \ref{sec-marl} reviews MARL algorithms (with particular focus on WPL and GIGA-WoLF, the two algorithms we use in our experimental evaluation). 
Section \ref{sec-results} presents our initial experimental results, where the global performance metric leads to a (misleading) conclusion that a MARL algorithm converges. 
Section \ref{sec-ied} presents our proposed measure and illustrates how it is used to expose the hidden instability of a MARL algorithm. 
We conclude in Section \ref{sec-conclude}.

\section{Case Study: Distributed Task Allocation Problem (DTAP)}
\label{sec-case}

\begin{figure}[htbp] \centering
\includegraphics[width=3.5in]{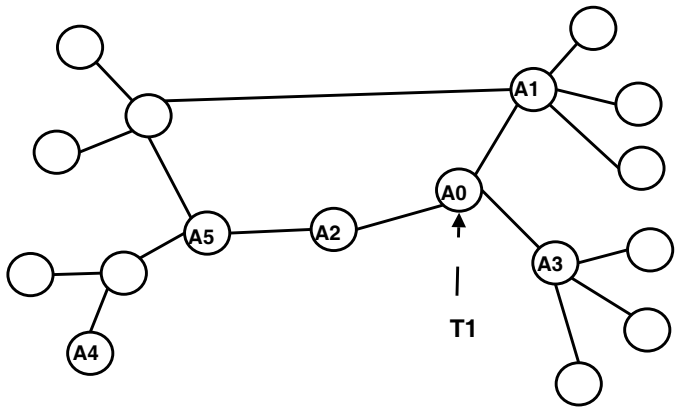}
      \caption{\small Task allocation using a network of agents.}
      \label{network2}
    \end{figure}

We use a simplified version of the distributed task allocation domain (DTAP) \cite{sherief08jair}, where the
goal of the system is to assign tasks to agents such that the
service time of each task is minimized. 
For illustration, consider the example scenario depicted in Figure
\ref{network2}. Agent A0 receives task T1, which can be executed by
any of the agents A0, A1, A2, A3, and A4. All agents other than
agent A4 are overloaded, and therefore the best option for agent A0 is
to forward task T1 to agent A2 which in turn forwards the task to its
left neighbor (A5) until task T1 reaches agent A4. Although agent A0 does
not know that A4 is under-loaded (because agent A0 interacts only with
its immediate neighbors), agent A0
should eventually learn (through experience and interaction
with its neighbors) that sending task T1 to agent A2 is the best action without
even knowing that agent A4 exists. 

The DTAP domain has an essential property that appears in many real world problems yet not captured by most of the domains that were used to analyze MARL algorithms experimentally: communication delay. The effect of an action does not appear immediately because it is communicated via messages and messages take time to route. Not only is the reward delayed but so is any change in the system's state. A consequence of communication delay is partial observability: an agent can not observe the full system state (the queues at every other agent, messages on links and in queues, etc.). 

Each time unit, agents make decisions regarding
all task requests received during this time unit.   
For each task, the agent can either 
execute the task locally or send the task to a neighboring agent. If
an agent decides to execute the task locally, the agent adds the task
to its local queue, where tasks are executed on a first come
first serve basis, with
unlimited queue length. 

 Each agent has a physical location. Communication delay between two agents
is proportional to the Euclidean distance between them, one time unit
per distance unit. 
Agents interact via two types of messages. A REQUEST message $\langle i, j, T
\rangle$ indicates a request sent from agent $i$ to agent $j$ requesting the execution of task $T$. An
UPDATE message $\langle i, j , T, R \rangle$ indicates a feedback (reward signal) from agent $i$ to agent $j$ that task $T$ took $R$ time steps to complete (the time steps are computed from the time agent $i$ received $T$'s request).

The main goal of DTAP is to reduce the total service time, averaged
    over tasks, $ATST = \frac{\sum_{T \in
    \overline{T}_{\tau}}TST(T)}{|\overline{T}_{\tau}|}$, where
$\overline{T}_{\tau}$ is the set of 
task requests received during a time period $\tau$ and $TST(T)$ is the
total time a task $T$ spends in the system. The $TST(T)$ time consists of the time for routing a task request 
through the network, the time the task request spends in the local
queue, and the time of actually executing the task. 

Although the underlying simulator has different underlying states, we deliberately made agents oblivious to these states. The only feedback an agent gets is its own reward. This simplifies the agent's decision problem and re-emphasises partial observability: agents collectively learn a joint policy that makes a good compromise over the different unobserved states (because the agents can not distinguish between these states).

\section{Multiagent Reinforcement Learning}
\label{sec-marl}
The experimental results in the following section focus on two gradient-ascent MARL algorithms: GIGA-WoLF \cite{bowling05nips} and WPL \cite{sherief08jair}.\footnote{A large number of MARL algorithms have been proposed that vary in their in their underlying assumptions and target domains \cite{panait05jaamas}. MARL algorithms that can only learn a deterministic policy (such as Q-learning \cite{sutton:rl}) are not suitable for the DTAP domain. For example, even if two neighbors have practically the same load, Q-learning will assign all incoming requests to one of the neighbors until a feedback is received later indicating a change in the load. On the other hand, an agent using a gradient ascent MARL algorithm has the ability to adjust its policy to a non-deterministic (or stochastic) distribution \cite{sherief08jair}. Q-learning was successfully used in the packet routing domain \cite{boyan94nips,dutta05jair}, where load balancing is not the main concern (the main objective is routing a packet from a particular source to a particular destination).} 
We chose these two algorithms because they allow agents to learn a stochastic policy based on the expected reward gradient. Both algorithms were also shown to converge in benchmark two-player-two-action games as well as some larger games.
The specifics of WPL and GIGA-WoLF (such as their update equations, the underlying intuition, their differences and similarities) are neither relevant to the purpose of this paper nor  needed to follow our analysis in Section \ref{sec-results}. 
Nevertheless, and for completeness, we mention below (very briefly) the equations for updating the policy for the two algorithms. 
Further details regarding the two algorithms can be found elsewhere \cite{bowling05nips,sherief08jair}. 

An agent $i$ using WPL updates its policy $\pi_i$ according to the following equations:

\begin{align*}
\forall j \in neighbors(i):\Delta\pi_{i}^{t+1}(j) & \leftarrow\frac{\partial V_{i}^t(\pi)}{\partial\pi_{i}^t(j)}\cdot\eta \cdot \left\{ \begin{array}{ll}
\pi_{i}^t(j) & \textrm{if }\frac{\partial V_{i}^t(\pi)}{\partial\pi_{i}^t(j)} < 0\\
1-\pi_{i}^t(j) & \textrm{otherwise}\end{array}\right.\\
\pi_{i}^{t+1} & \leftarrow projection(\pi_{i}^t+\Delta\pi_{i}^{t+1})\end{align*}

where $\eta$ is a small learning constant and $V_i(\pi_i)$ is the expected reward agent $i$ would get if it interacts with its neighbors according to policy $\pi_i$. The $projection$ function ensures that after adding the gradient $\Delta \pi_i$ to the policy, the resulting policy is still valid.

An agent $i$ using GIGA-WoLF updates its policy $\pi_i$ according to the following equations:

\begin{align*}
\hat{\pi}_i^{t+1} & = projection(\pi_i^t + \eta V_i^t(\pi)^t) \\
z_i^{t+1} & = projection(\pi_i^t + \eta V_i^t(\pi) / 3) \\
\delta_i^{t+1} & = \min \left( 1 , \frac{||z_i^{t+1} - z_i^t||}{z_i^{t+1} - \hat{\pi}_i^t} \right) \\
\pi_i^{t+1} & = \hat{\pi}_i^{t+1} + \delta_i^{t+1} (z_i^{t+1} - \hat{\pi}_i^{t+1})
\end{align*}

\section{Stability Under the Global Metric Results}
\label{sec-results}
\begin{figure}[htbp] \centering
\includegraphics[width=3.5in]{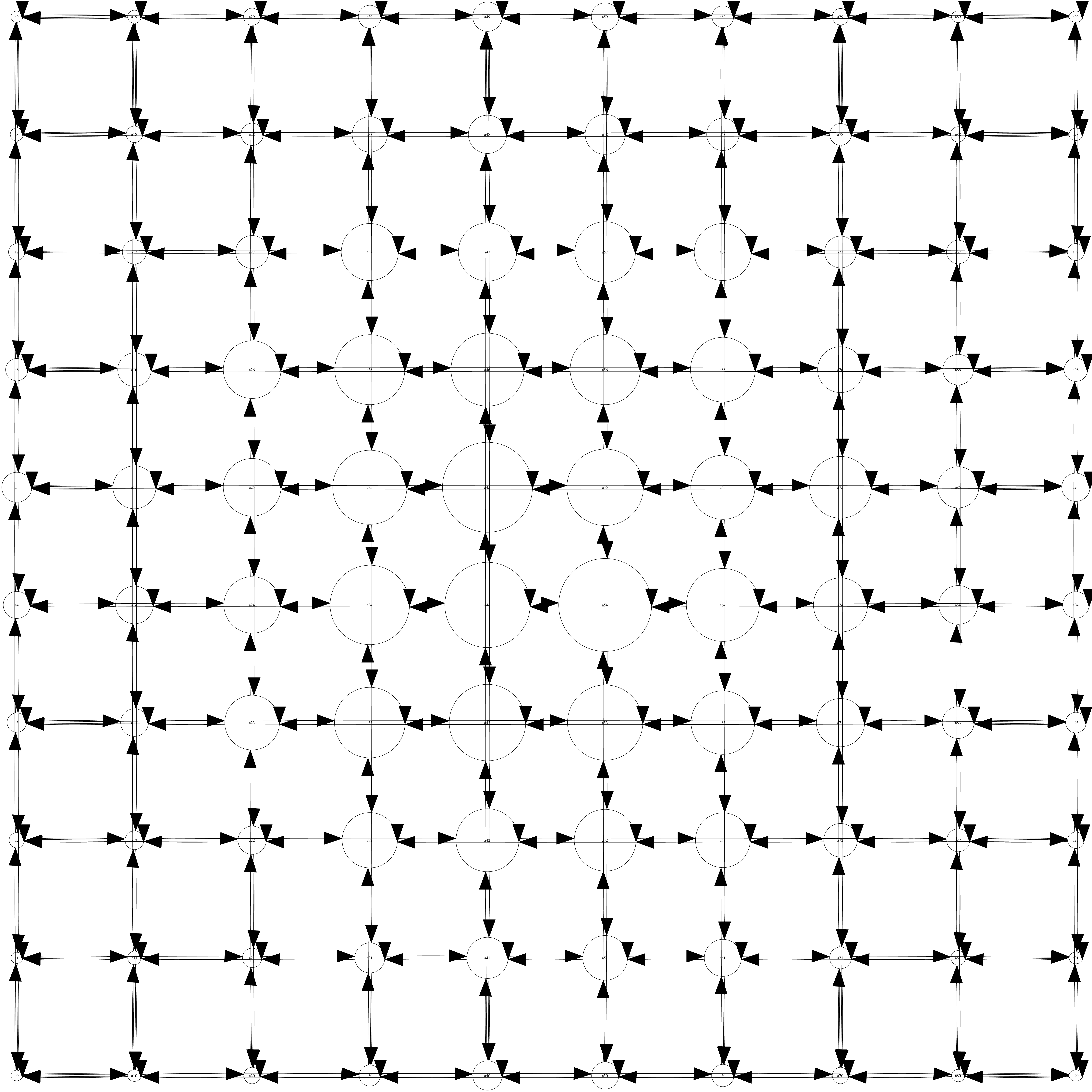}
      \caption{\small The simulation setting for the DTAP domain. Only the 16 nodes at the center receive tasks from the environment. A node's diameter reflects its local queue length.}
      \label{fig-setting}
    \end{figure}

We have evaluated the performance of WPL and GIGA WoLF using the following setting.\footnote{The simulator is available online at \url{http://www.cs.umass.edu/~shario/dtap.html}.} 100 agents are organized in a 10x10 grid. Communication delay between two adjacent agents is two time units. Tasks arrive at the 4x4 sub-grid at the center at rate 0.5 tasks/time unit. All agents can execute a task with a rate of 0.1 task/time unit (both task arrival and service durations follow an exponential distribution). Figure \ref{fig-setting} illustrates the setting.

\begin{figure}[htbp] \centering
\includegraphics[width=5in]{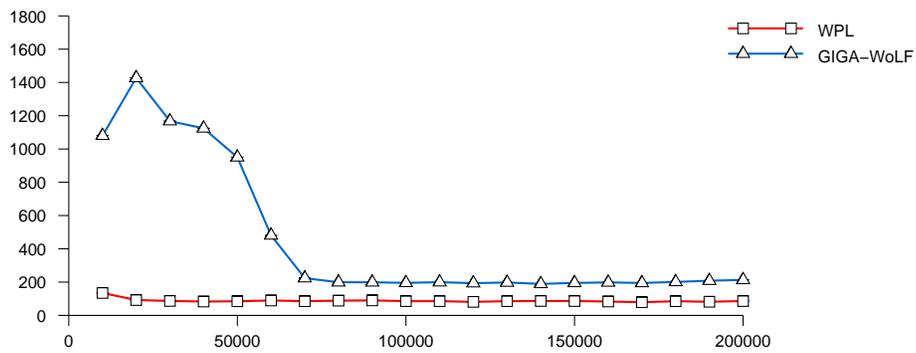}
      \caption{\small Comparing the average total service time for 200,000 time steps of the DTAP problem for WPL and GIGA-WoLF.}
      \label{fig-performance}
    \end{figure}

Figure \ref{fig-performance} plots the global performance (measured in terms of ATST) of the two multi-agent learning algorithms in the DTAP domain. Just by looking at ATST plot, it is relatively safe to conclude that WPL converges quickly while GIGA-WoLF converges after about 75,000 time steps. The following section presents the measure we have used in order to discover that the stability of GIGA-WoLF is actually spurious.

\section{Verifying Stability Using Policy Entropy}
\label{sec-ied}

Ideally, we would want to visualize and analyze the evolution of all learning parameters, including action values and the policy of all agents. However, going to such detail is only possible for small number of agents. As the number of agents increases, one needs aggregated measures that summarize the system's behavior yet can reflect the stability of the system's dynamics. 

We propose using a simple measure that summarizes an agent's policy into a single number: the policy entropy, $H(\pi_x)$, for a particular agent $x$:

\[
H(\pi_x) = -\sum_{y \in neighbors(x)} \pi_x(y) lg \pi_x(y)
\] 

The function $\pi_x$ is the policy of agent $x$ ($\pi_x(y)$ is the probability that agent $x$ interacts with its neighbor $y$). In case of gradient ascent, the policy is explicitely learned. For deterministic learners (such as Q-learning), the effective policy can be estimated by counting the number of times each neighbor is chosen. 

\begin{figure}[htbp] \centering
\includegraphics[width=4in]{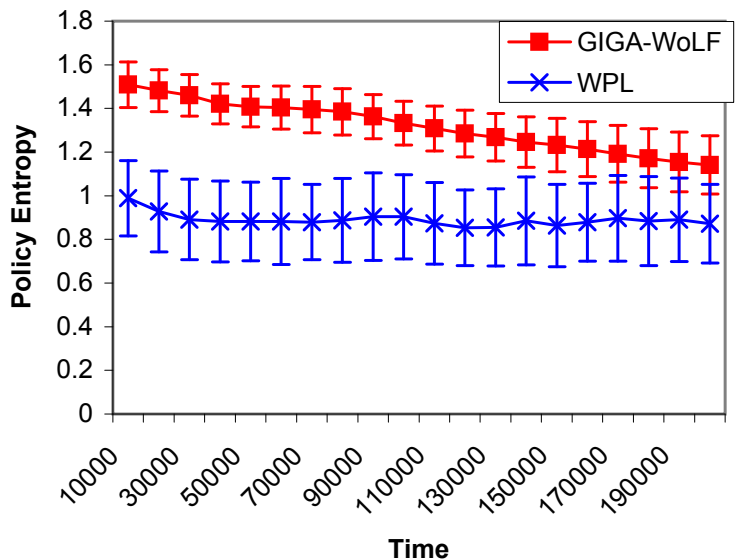}
      \caption{\small The policy entropy of WPL and GIGA-WoLF for 200,000 time steps.}
      \label{fig-ent-short}
    \end{figure}

Figure \ref{fig-ent-short} plots the average policy entropy and the associated standard deviation, over the 100 agents, against time. Agent policies under WPL do converge but the policies under GIGA-WoLF have not converged yet. The policy entropy is still decreasing, which suggests that GIGA-WoLF is still adapting.

This intrigued us to rerun the simulator, this time allowing the simulator to run for 600,000 times steps instead of just 200,000 time steps. To our surprise, the global performance metric (the ATST in this case) of GIGA-WoLF starts slowly to diverge after 250,000 time steps and the corresponding policy entropy continue to decrease. WPL's policy entropy remains stable, as well as the global performance metric.

    \begin{figure}[htbp] \centering
\includegraphics[width=5in]{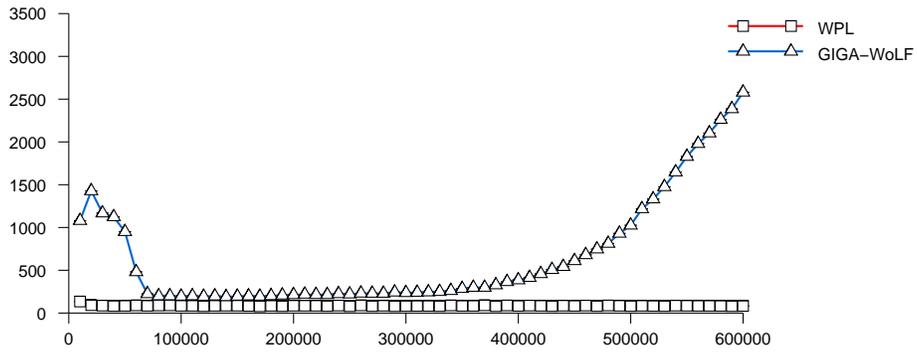}
      \caption{\small Comparing the average total service time for 600,000 time steps of the DTAP problem for WPL and GIGA-WoLF.}
      \label{fig-performance-long}
    \end{figure}

    \begin{figure}[htbp] \centering
\includegraphics[width=4in]{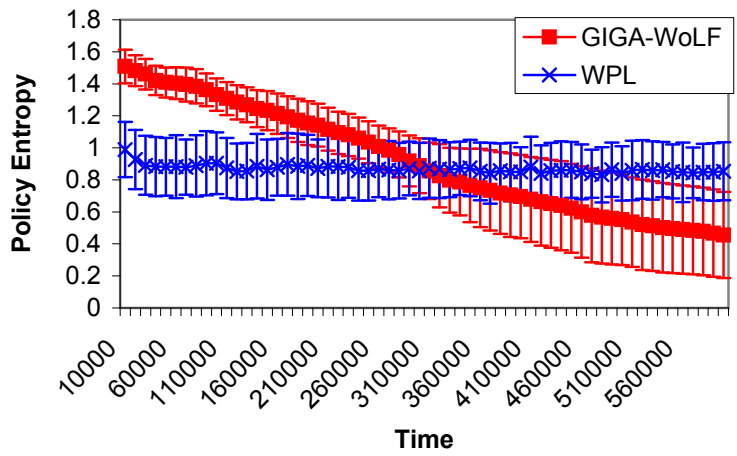}
      \caption{\small The average policy entropy of WPL and GIGA-WoLF for 600,000 time steps.}
      \label{fig-ent-long}
    \end{figure}

More in-depth analysis is needed in order to fully understand the dynamics of GIGA-WoLF and WPL in the DTAP domain and in large-scale systems in general. However, this is beyond the scope of this research note and should not distract from the main point we are trying to make: the common practice of using a global performance metric to  verify the stability of a MARL algorithm is not reliable and can be misleading.

\section{Conclusion and Future work}
\label{sec-conclude}

The main contribution of this paper is showing that using a global performance metric for verifying the stability (or even the usefulness) of a MARL algorithm is not a reliable methodology. In particular, we present a case study of 100 agents where the global performance metric can hide an underlying instability in the system that later leads to a significant drop in performance. We propose a measure that successfully exposes such instability.

One of the issues indirectly raised by this paper is for how long shall a performance metric be stable in order to conclude the stability of the underlying MARL algorithm? 
Currently, no theoretical framework addresses this question, which we believe to be an essential requirement for adopting MARL in practical large-scale applications.

%------------------------------------------------------------------------- 
%\nocite{ex1,ex2}
\bibliographystyle{elsarticle-num}
\bibliography{references}

\end{document}